%% file: main.tex
\documentclass[twocolumn]{aastex62} 

\include{definitions}

\usepackage{multirow}

\received{XXXX}
\revised{XXXX}
\accepted{XXXX}
\submitjournal{ApJ}

\shorttitle{Fermi LAT observations of \sgra}
\shortauthors{Cafardo \& Nemmen}

\begin{document}

\title{\textit{Fermi} LAT observations of Sagittarius A*: Imaging Analysis}

\correspondingauthor{Fabio Cafardo} 
\email{fabio.cafardo@usp.br}

\author{Fabio Cafardo \orcid{https://orcid.org/0000-0002-7910-2282}}
\affil{Universidade de S\~ao Paulo, Instituto de Astronomia, Geof\'{\i}sica e Ci\^encias Atmosf\'ericas, Departamento de Astronomia, S\~ao Paulo, SP 05508-090, Brazil}

\author{Rodrigo Nemmen \orcid{https://orcid.org/0000-0003-3956-0331}}
\affil{Universidade de S\~ao Paulo, Instituto de Astronomia, Geof\'{\i}sica e Ci\^encias Atmosf\'ericas, Departamento de Astronomia, S\~ao Paulo, SP 05508-090, Brazil}
\collaboration{(Fermi LAT Collaboration)}

\begin{abstract}
Sagittarius A* (\sgra)---the supermassive black hole (SMBH) in the center of our galaxy---has been observed in most of the electromagnetic spectrum, from radio to X-rays. Diffuse $\gamma$-ray emission has been observed around \sgra\ and a $\gamma$-ray point source has been detected coinciding with the SMBH's position, although there is no definitive association between the two. In this work, we have used $\sim$11 years of \fermi\ Large Area Telescope (LAT) observations of the point source 4FGL J1745.6$-$2859 and performed a detailed imaging analysis across four energy bands. Our goal is to elucidate the nature of the $\gamma$-ray emission at the Galactic Center (GC) and whether it is associated with the SMBH. We find that the centroid of the emission approaches \sgra's location as the energy increases. Assuming that the $\gamma$-ray point source is located at the GC, we estimate a luminosity of $2.61 \times 10^{36} \ {\rm erg \ s}^{-1}$ in the 100 MeV to 500 GeV energy range. This is consistent with \sgra's bolometric luminosity. Based on the point source properties, we ruled out several potential candidates for its nature and favor a cosmic ray origin either from protons, electrons or both, accelerated by---or in the vicinity of---the SMBH. Our results indicate that the point source at the GC is indeed the $\gamma$-ray counterpart of \sgra\ in the GeV range.
\end{abstract}


\section{Introduction} \label{sec:intro}

The center of our galaxy hosts a SMBH with a mass of $\sim$ $10^6 M_\odot$ \citep{Ghez08, Genzel10, Boehle16} located at a distance of 8.2 kpc \citep{Gravity2019}. Like the Milky Way, it is believed that every sufficiently massive galaxy harbors a SMBH in its center \citep{Lynden-Bell69, Kormendy95, Miyoshi95, Heckman14}. The first observations of a source that would later be associated with \sgra---the SMBH in our GC---were made by \cite{Balick74} in radio wavelengths. Almost fifty years later, there are several observations that confirm the presence of the SMBH. The most convincing are those obtained by \cite{Eisenhauer05} and \cite{Gillessen09} monitoring stellar orbits and the detection of orbital motions of a ``hot spot'' in the accretion flow near the last stable circular orbit of \sgra\ \citep{GravityCollab18}. Results of the Event Horizon Telescope observational campaign to reveal the image of the shadow of \sgra\ on the accretion flow are still anxiously anticipated \eg{EHTC2019}.

The electromagnetic radiation from \sgra\ has been seen in several wavelengths \eg{Genzel10, Morris2012, Eckart18}. It is highly variable in the infrared \citep{Genzel03, Ghez04, Hornstein07, Dodds-Eden09, Witzel12, Hora14, Witzel18, von_Fellenberg18, Fazio18, Boyce19} and X-rays \citep{Baganoff01, Nowak12, Neilsen13, Barriere14, Neilsen15, Ponti15, Fazio18, Boyce19}, which suggests a compact source. Variability has also been reported in longer wavelengths \citep{Zhao03, Miyazaki04, Mauerhan05, Macquart06, Yusef-Zadeh06, Marrone08, Yusef-Zadeh09, Dexter14b, Brinkerink15, Stone16}.

Following the H.E.S.S. TeV detections of the GC \citep{Aharonian04}, several candidates have been proposed for this $\gamma$-ray flux: \sgra\ itself, either from its immediate vicinity \citep{Aharonian05} or from a ``plerion'' produced by the SMBH winds \citep{Atoyan04, Kusunose12}; the interaction between the dense molecular clouds in the GC with cosmic rays accelerated by \sgra\ and/or by some other nearby source \citep{Aharonian&Neronov05, Ballantyne11, Chernyakova11, Linden12, Fatuzzo12, Guo13}; the pulsar wind nebula (PWN) G359.95-0.04 \citep{Wang06, Hinton07}; the supernova remnant (SNR) Sagittarius A East \citep{Crocker05} (but see \citealt{Aharonian09, Acero10}); self-annihilating dark matter particles accumulating at the GC \citep{Hooper11, Hooper&Linden11} and an as-yet undetected pulsar (or population of pulsars) \citep{Hooper&Linden11}. The GC $\gamma$-ray flux does not seem to be variable \citep{Chernyakova11, Malyshev15, Ahnen17}.

It is believed that the quiescent state of \sgra, observed from radio to X-rays, is due to a radiatively inefficient accretion flow (RIAF) \eg{Narayan95nat, Yuan14}. The broadband spectrum is dominated by the radio-to-submm emission which is understood as synchrotron radiation from a thermal population of electrons, with temperatures between $\sim$5$-$20 MeV, as well as a small fraction (a few percent) of nonthermal electrons \citep{Yuan03}. In its steady state, \sgra\ emits $\sim$10$^{36} \ {\rm erg \ s}^{-1}$ \eg{Genzel10}. There is evidence that the X-ray flare emission is due to synchrotron processes \citep{Dodds-Eden09, Barriere14, Ponti17} although they have also been interpreted as Inverse Compton (IC) upscattered photons by the mildly relativistic, nonthermal electrons \citep{Yusef-Zadeh09, Ball2016}.

Prominent gamma-ray emission from MeV to TeV energies coincident with \sgra's position is observed by \fermi-LAT. Since the beginning of \fermi's operations, a point source has been observed coinciding with the position of \sgra. This source was studied by \cite{Chernyakova11} with 25 months of \fermi\ observations. They found no temporal variability at GeV energies and proposed a model in which the $\gamma$-ray emission in the inner 10 pc of the Galaxy arises from relativistic proton interactions. Later, \cite{Malyshev15} analyzed the same source using 74 months of data and the Second Catalog of \fermi-LAT Sources \citep{2fgl}. They also found no variability in the flux, and considered the observed spectrum as consistent with IC scattering of high-energy electrons.

\cite{Ahnen17} collected several models for the MeV to TeV emission from \sgra. The list includes leptonic \citep{Kusunose12}, hadronic \citep{Fatuzzo12, Linden12, Ballantyne11, Chernyakova11} and hybrid \citep{Guo13} models. We call the models in this list ``Fermi-era'', since they were all constructed taking in consideration \fermi-LAT's data.

The TeV emission observed by H.E.S.S. indicates the presence of PeV protons within the central tens of pc of the Galaxy \citep{HESSCollaboration16}. They propose that a more active phase of \sgra\ in the past could have accelerated this population of high-energy protons. 
There is also tantalizing evidence for an enhanced level of activity in the recent past of \sgra\ through the \fermi\ Bubbles \citep{Su10} which should have formed 1--3 Myr ago and endured for 0.1--0.5 Myr \citep{Guo12, Yang18}. The origin of the bubbles is still debated and could be also due to a previous starburst in addition to the activity of the SMBH. Further evidence for higher levels of activity in \sgra\ comes from X-ray observations of circumnuclear clouds \citep{Ponti10}. Concretely, X-ray observations since the 1990s show rapid variations in the 6.4 keV of Fe K$\alpha$ line propagating through molecular clouds in the inner Galactic regions. These variations are likely the result of a highly variable active phase of \sgra\ within the past few hundred years, which is echoing through the clouds. Models indicate at least two luminous outbursts ($\sim$100 and 400 years ago) on few-year timescales during which the luminosity of \sgra\ went up to at least 10$^{39}$ erg s$^{-1}$ \citep{Ponti10,Clavel13}. In summary, it seems that \sgra\ was 10$^3$ times more active within the past few centuries compared to current levels. 

The GC is the closest example of a galactic nucleus and a compelling laboratory to investigate the physical processes responsible for accelerating particles to TeV and PeV energies. Of the several sources near the GC in the Fourth catalog of \fermi-LAT sources (4FGL, \citealt{4FGL}), 4FGL J1745.6$-$2859 is the brightest and the closest to \sgra, at a distance of $\sim0.01^{\circ}$. Furthermore, it has been shown that low-luminosity active galactic nuclei, of which \sgra\ is an example, can be efficient $\gamma$-ray emitters \citep{de_Menezes20}. Hence, it is logical to propose \sgra\ as a contributor to the GC $\gamma$-ray flux. Here, we report an imaging analysis for this point source. In section  \ref{sec:obs} we describe the observations and data analysis procedure. In section \ref{sec:res} we describe the results from the imaging analysis for the point source in four different energy ranges between 60 MeV--500 GeV. In section \ref{sec:disc} we discuss the possible candidates that can explain the observed $\gamma$-ray emission. Finally, in section \ref{sec:summary} we summarize our main findings.

\section{Observations and Data Analysis} \label{sec:obs}

In this work, we divided the analysis into four energy bands: 60--300 MeV, 300 MeV--3 GeV, 3--10 GeV and 10--500 GeV. The goals were to study the impact of photon energy on the location of the source, to measure 4FGL J1745.6$-$2859's $\gamma$-ray emission in different parts of the electromagnetic spectrum to allow comparisons with predictions of several models that try to explain the GC's $\gamma$-ray flux (Section \ref{sec:disc}) and also to take advantage of the better \fermi-LAT  point-spread function (PSF) at higher energies.

\begin{figure}
\centering
\includegraphics[width=\linewidth]{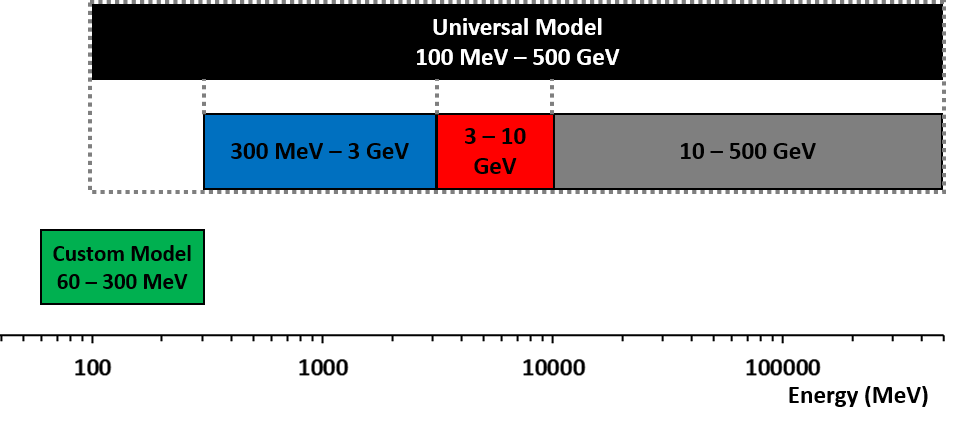}
\caption{We divided our analyses into four energy bands (60--300 MeV, 300 MeV--3 GeV, 3--10 GeV and 10--500 GeV). The models for the three highest-energy bands were created based on an analysis that considered energies in the range 100 MeV--500 GeV. The lowest-energy band was studied with a custom model.}
\label{fig:Models}
\end{figure}

The investigations of the three highest-energy bands are based on an analysis (which we call ``universal model'') performed with energies between 100 MeV and 500 GeV. The lowest-energy band used a custom model. Figure \ref{fig:Models} shows how we split the analysis into different energy bands and models.

This study has some similarities to the work of \cite{Malyshev15}, with several improvements. The most obvious is the longer time baseline of the photon data---more than 10.5 years of observations versus 6 years---which provides better statistics. This allows for sharper images and better modelling with reduced source confusion. Also, we are using updated versions of the analysis tools and a more-recent version of the \fermi\ catalog (accompanied with improved versions of the diffuse models), all of which were released after their work. Finally, we chose to use a stricter \textit{event type} selection than \cite{Malyshev15}.

In this Section we show how the universal model (and its descendants) and the low-energy custom model were created and how they were used to assess 4FGL J1745.6$-$2859's $\gamma$-ray flux and position in different energies.

\subsection{The universal model and its descendants} \label{sec:obsHE}

Here we describe the process for the creation of a spatial/spectral model with photons between 100 MeV and 500 GeV. This model was used to evaluate the $\gamma$-ray photon and energy fluxes of the source. Also, it was used as the basis for three different analyses considering only photons with energies between 300 MeV--3 GeV, 3--10 GeV and 10--500 GeV.

In this part of the work, we used $\sim$11.3 years (from 2008 August to 2019 December) of \fermi-LAT \citep{lat09} data.  We considered a region of $20^{\circ} \times 20^{\circ}$ square centered on the point source 4FGL J1745.6$-$2859 coincident with \sgra\ and rotated $58{^\circ}.6$ to the East in Galactic coordinates. In this work, we use photons classified as \code{SOURCE}.

\fermi-LAT photons are classified in different \textit{event types}. They can be categorized according to the location where they converted in the detector (photons converted in the \code{front} of the equipment have better-measured directions than those converted in the \code{back}) or according to the quality of the reconstructed direction (the photons are divided in quartiles depending on the quality of this reconstruction). In this work we chose to consider only the 75$\%$ photons with the best-reconstructed directions (\textit{event types} \code{PSF1}, \code{PSF2} and \code{PSF3}).

Data were binned to a pixel size of 0$^{\circ}$.08. We chose the recommended\footnote{\begin{scriptsize}https://fermi.gsfc.nasa.gov/ssc/data/analysis/documentation/\\Cicerone/Cicerone$\_$Data$\_$Exploration/Data$\_$preparation.html\end{scriptsize}} value of $\geqslant 90^{\circ}$ for the zenith angle cut.

The region was modeled based on the sources' positions and spectral models in the preliminary release of 4FGL (\code{gll\_psc\_v20.fit}, \citealt{4FGL}), the updated model of interstellar $\gamma$-ray emission, \code{gll\_iem\_v07.fits}, and standard isotropic spectral templates selected according to the \textit{event types} and \textit{event class} used in this work. We performed a binned likelihood analysis using Fermitools conda package version 1.2.1, \code{Fermipy} \code{python} package version 0.17.4 \citep{wood2017fermipy} and Pass 8 release 3 Version 2 response functions \citep{atwood2013pass}. Energy dispersion was disabled for the isotropic diffuse component only.

To perform a likelihood analysis with \fermi-LAT data, it is necessary to create a model of the $\gamma$-ray emitting sources in the region of interest (RoI). To generate the model, we included the 4FGL sources inside a square larger than the size of the RoI (with $25^{\circ}$ side) to account for \fermi-LAT's PSF. Its PSF varies considerably with photon energy. It ranges from  $\lesssim$10$^{\circ}$ (68$\%$ containment for the photon selection used in this work) for photons at 60 MeV to $\lesssim$0$^{\circ}$.2 for $\gtrsim$10 GeV photons.

We performed a joint likelihood analysis with three components accounting for the isotropic emission because \textit{event types} \code{PSF1}, \code{PSF2} and \code{PSF3} have different isotropic spectral templates. We started by creating the universal model with energies between 100 MeV and 500 GeV. In the 4FGL Catalog, the 4FGL J1745.6$-$2859 spectral model is a log-parabola:
\begin{equation}
    \frac{dN}{dE}=N_0\left(\frac{E}{E_0} \right)^{-\alpha-\beta \log(E/E_0)}
\label{Eq:LogNorm}
\end{equation}
where $dN/dE$ is the differential photon flux, $N_0$ is the normalization, $E_0$ is a scale parameter, $\alpha$ is the spectral slope at $E_0$ and $\beta$ gives the curvature of the spectrum.

We began the analysis using \code{Fermipy}'s \code{optimize} method. This tool fits the spectral models of all sources within the RoI through an iterative strategy: it starts by simultaneously fitting  the normalization parameters of the brightest sources, then individually fits the normalization of every source not included in the first step, and finishes by individually fitting the normalizations and indexes of sources with the largest TS. After that we started fitting the sources in the RoI. Since the GC region is very rich in sources---there are 196 4FGL sources in the central $10^{\circ}$ of the Galaxy---we used an iterative approach for fitting of the RoI, always taking care to limit the number of free parameters to $\lesssim 15$ as recommended by the optimizer's (MINUIT) manual. Several iterations were performed in order to fit the brightest sources closest to 4FGL J1745.6$-$2859. In all iterations, the normalization of our source of interest was allowed to vary as well as the normalizations of the Galactic diffuse emission model (\code{galdiff}) and of the isotropic spectral template (\code{isodiff}). The normalizations of selected sources were also iteratively freed based on their proximity to the center of the RoI and their brightness in $\gamma$-rays, measured by the number of predicted photons in each energy interval. We performed several iterations to fit the desired sources. Only the best-quality fits (Quality: 3, Status: 0) were considered. Fits that did not converge or with lower quality were disregarded. In this case we would go back to the previous step of the fitting procedure and continue from there with fewer free sources.

The next step was to use the \code{Fermipy} function \code{find\_sources}. Twenty-nine new sources were found with this tool in the energy range between 100 MeV and 500 GeV. We used power-law spectral models for each of the new sources. This spectral model was selected because it is typically adequate for relatively faint sources, and in fact the majority of sources in 4FGL are modeled with a power-law spectrum. Finding and characterizing new sources is not a goal of this work. The aim of this source-finding step was to improve the quality of the model. Some of the newly found sources are likely spurious detection due to unmodeled background emission. The new sources are listed in Appendix \ref{sec:newsources}.

In between these steps we evaluated the quality of our fitting procedures by two approaches:
\begin{itemize}
\item Residual maps: built by subtracting the modeled counts from the data and searching for regions with significant residuals\footnote{For a description on how the significance is calculated:\\ https://fermipy.readthedocs.io/en/latest/advanced/residmap.html}.
\item Test Statistic (TS, \citealt{Mattox96}) maps: searching for the presence of an additional source component in each spatial bin of the RoI.
\end{itemize}

Maps that show no regions with TS $\gtrsim$ 25 and Residuals $\gtrsim |4\sigma|$ were considered well modeled. Regions with excesses above these levels were fit again. In these cases, the normalizations of sources close to excesses in those maps were allowed to vary, together with the normalizations of 4FGL J1745.6$-$2859 and of the diffuse components. Usually, mainly in the case of negative residuals, this additional round of fitting was enough to lessen the excesses in the maps while also increasing the likelihood of the model. In the case of regions with positive TS, some of the excesses were reduced only after using the \code{find\_sources} tool.

The last step was fitting the spectral indexes of the central source. The approach was the same as described above. We performed different iterations of fitting, always with the normalization and the indexes of 4FGL J1745.6$-$2859 as free parameters together with other free parameters that usually included the normalization and the spectral indexes of nearby bright sources and the Galactic and isotropic diffuse models. Only iterations with the best fit quality were considered.

These steps led to the creation of the universal model with photons between 100 MeV and 500 GeV. Then, it was used as an initial model for the analysis in three energy bands (300 MeV--3 GeV, 3--10 GeV and 10--500 GeV). In each energy band, the only sources that were allowed to vary in a new round of fitting were 4FGL J1745.6$-$2859 (normalization and index) and the Galactic and isotropic diffuse models (normalization only). This ``minimal fitting'' was used with the objective of keeping a similar model in the three energy bands.

To determine the centroid of the $\gamma$-ray emission of the central point source for each energy range and calculate the likelihood of it being spatially extended, we used \code{Fermipy}'s \code{extension} method. In addition to finding the location of the point source, it computes the likelihood of the source being extended with respect to it being pointlike. Also, it gives the best-fit model for extension. We chose to use a 2D Gaussian as the putative extended spatial model for the central source.

\subsection{The low-energy model} \label{sec:obsLE}

Using the universal model as a starting point for a new model at lower energies proved to be challenging. We could not obtain a good-quality model with the same minimal fitting used to ``split'' the universal model in 3 energy bands (described at the end of Section \ref{sec:obsHE}) nor with several additional rounds of fitting: in both cases we ended up with models whose Residuals and TS maps showed many regions with excesses above the acceptable levels. So we decided to use a model created specifically for the 60--300 MeV energy band. We already had this model prepared from previous studies. As will be described below, some features of this model's data are different from the universal model's. This is not a serious issue because, in this work, we treat the results for each energy band independently, avoiding comparisons between each other.

We considered data inside a $30^{\circ} \times 30^{\circ}$ square also centered on the point source 4FGL J1745.6$-$2859 and with the same orientation as the one used to the universal model. This choice of size for the RoI is appropriate for modeling the distribution of lower-energy photons given the broad PSF. We used 0$^{\circ}$.1 pixel size and the recommended value of $\geqslant 90^{\circ}$ for the zenith angle cut. About 10.5 years of \fermi-LAT data was considered (from 2008 August to 2019 February).

We modeled the region using the preliminary release of 4FGL (\code{gll\_psc\_v17.fit}, \citealt{4FGL}), the updated model of interstellar $\gamma$-ray emission, \code{gll\_iem\_v07.fits}, and standard isotropic spectral templates. We included in the model all 4FGL sources in a region with $35^{\circ}$ side to account for \fermi-LAT's PSF. Energy dispersion was disabled for the isotropic and Galactic diffuse components.

As for the analysis described in Section \ref{sec:obsHE} we used \textit{event class} \code{SOURCE}, \textit{event types} \code{PSF1}, \code{PSF2} and \code{PSF3} and performed a joint likelihood analysis with three components accounting for the isotropic emission.

Before starting the analysis we changed the Spectrum Type of 4FGL J1745.6$-$2859. It is cataloged as log-parabola but we adopted a power-law in this low-energy model (Equation \ref{Eq:PL}, where $\Gamma$ is the spectral slope.). We used results obtained in previous analysis using the Third Fermi LAT catalog (3FGL, \citealt{3fgl}) as the starting values for the parameters that were later refit with the new data. The main reason for the change was the ease of fitting power-law spectra, which have one less parameter compared to log-parabolas. This change is appropriate since in narrow energy bands (like the ones we are using here) a log-parabola can be approximated by a power law:
\begin{equation}
    \frac{dN}{dE}=N_0\left(\frac{E}{E_0} \right)^{-\Gamma}
\label{Eq:PL}
\end{equation}

After this, we followed the same process described in Section \ref{sec:obsHE} to fit the model. During this process, we found 14 new sources with \code{Fermipy}'s function \code{find\_sources} in regions associated with TS $>$ 25. But in new rounds of fitting, several of them showed a reduction of their TS to values below 25. Since \fermi-LAT's PSF is broader at lower energies, we decided to exclude---one at a time---the new sources in this condition from the model to avoid them interfering in the results of our source of interest. We started by excluding the farthest from the center of the RoI and refit the model. The normalizations of the sources closest to the excluded one were left free, together with the normalizations of the Galactic diffuse emission model, of the isotropic spectral template and of 4FGL J1745.6$-$2859. We repeated this process until there were no new sources with TS $<$ 25 in the model. After excluding these sources, we ended up with 5 new sources. They are listed in Appendix \ref{sec:newsorcesLowEnergy}. As we mentioned in Section \ref{sec:obsHE}, these new sources may be spurious detections due to inaccuracies in the models. Finally, we used \code{Fermipy}'s \code{extension} method to assess the location of 4FGL J1745.6-2859 in this energy range.

\section{Results}  \label{sec:res}

\begin{figure}
\centering
\includegraphics[width=\linewidth]{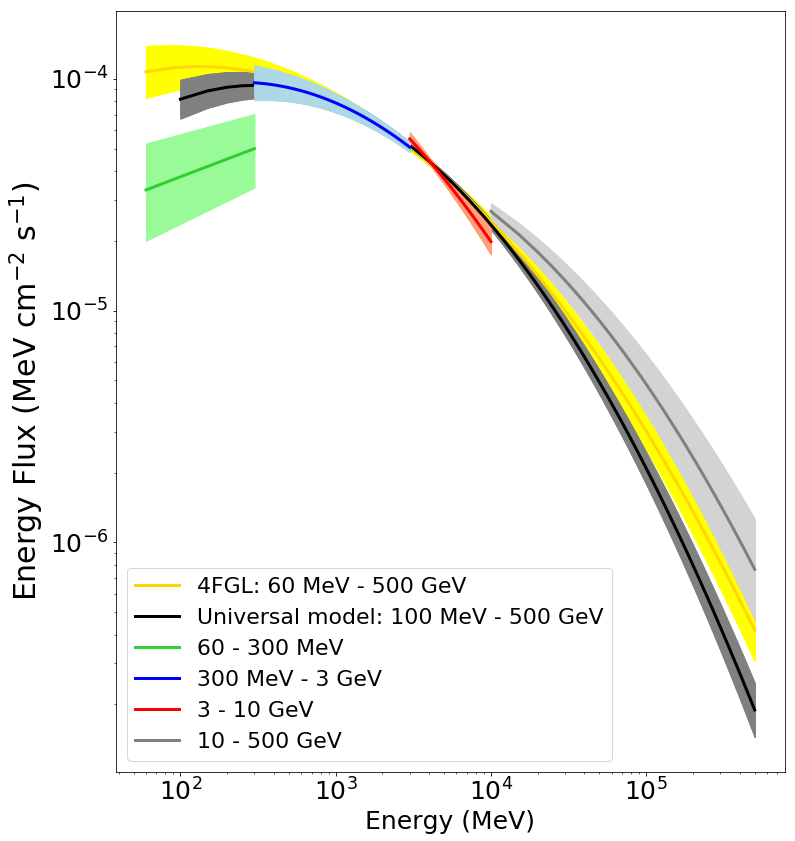}
\caption{Comparison of spectral models for 4FGL J1745.6$-$2859. The black line shows the universal model created in the 100 MeV--500 GeV energy range. It was later used as basis to create models in three different energy bands: 300 MeV--3 GeV (blue), 3--10 GeV (red) and 10--500 GeV (gray). They were created with log-parabolas templates. A power-law model was used in the 60--300 MeV energy range (green). The log-parabola spectral model used by the 4FGL Catalog for this source is also shown (yellow).}
\label{fig:spectra}
\end{figure}

In our work, we subdivided the analysis into four energy ranges. In Figure \ref{fig:spectra} we compare the results of these four models with the log-parabola spectral model adopted in the 4FGL Catalog for 4FGL J1745.6$-$2859. We can see there is a considerable discrepancy between the low (60--300 MeV) and higher energy ($>$300 MeV) spectral models. This is the result of the different modelling for the 60--300 MeV band, as discussed in Section \ref{sec:obs}. Only part of this difference can be explained by the addition in our 60--300 MeV model of a new source (not included in 4FGL but listed as PS J1750.6-2723 in Appendix \ref{sec:newsorcesLowEnergy}) at a distance $<$2$^{\circ}$ from 4FGL J1745.6$-$2859. This difference means that the $\gamma$-ray flux for this energy band might be underestimated. It is important to notice that the GC is among the most complicated regions in the sky to study with \fermi\ data: in addition to the high density of sources, the region is also engulfed by the Galactic diffuse emission. These factors are enhanced in lower energies due to the large PSF.

\begin{figure*}
 \centering
 \begin{tabular}{@{}c@{}}
  \includegraphics[width=.49\linewidth]{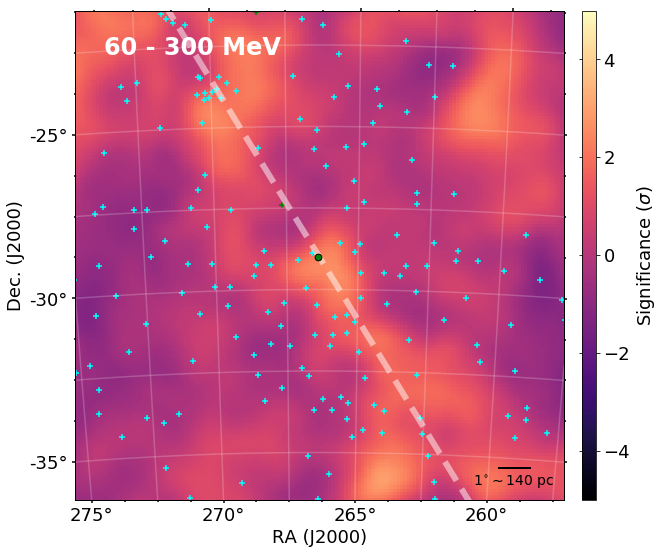}
  \hspace{0.02\linewidth}
  \includegraphics[width=.49\linewidth]{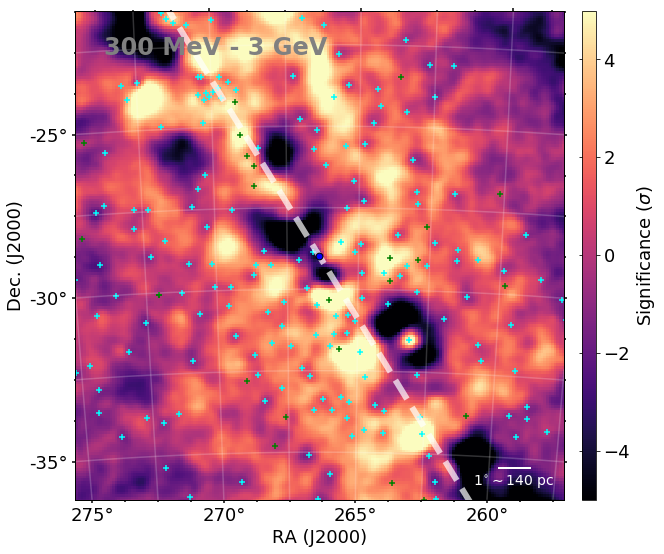} 
 \end{tabular} \\ 
 \begin{tabular}{@{}c@{}}
  \includegraphics[width=.49\linewidth]{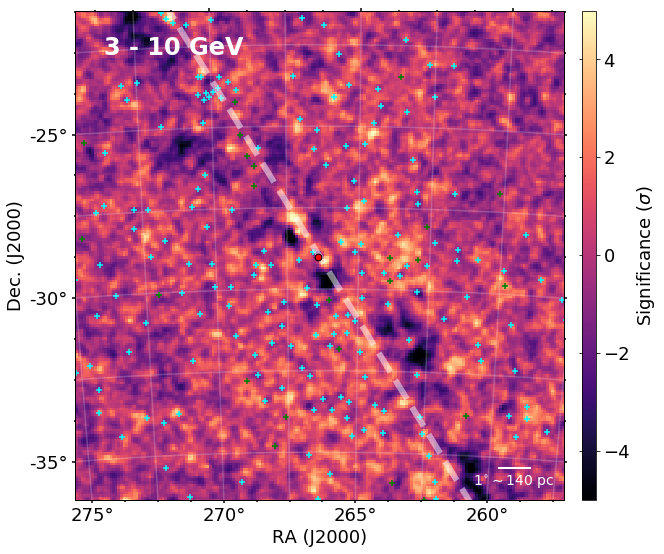}
  \hspace{0.02\linewidth}
  \includegraphics[width=.49\linewidth]{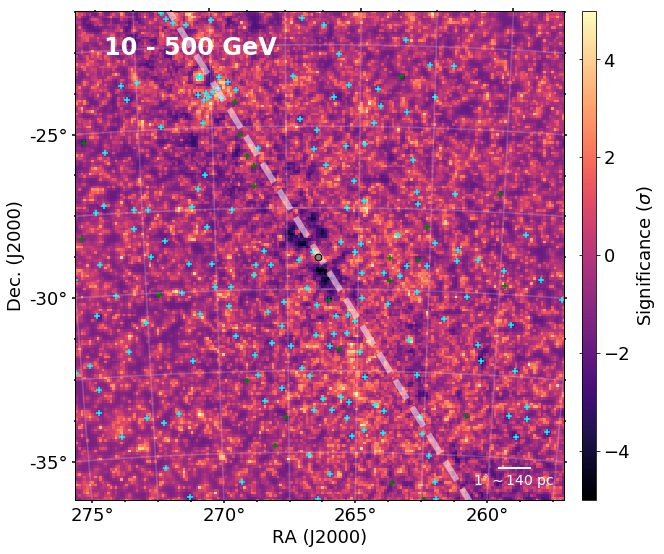}
 \end{tabular}
 \caption{Residual maps for the four different energy ranges. The colors show the significance of the residual. The point at the center of each panel corresponds to the central source position obtained in each energy range. 4FGL point sources are displayed as cyan crosses and new sources found during the analysis as green crosses. The gray dashed lines indicate the direction of the Galactic equator. An angular separation of 1$^{\circ}$ corresponds to $\sim$ 140 pc at \sgra's distance (8.2 kpc).}
\label{fig:Resid}
\end{figure*}

In Table \ref{TableSpectralParameters} we show the spectral parameters for 4FGL J1745.6$-$2859 obtained from the fitting process detailed in Section \ref{sec:obs} and shown in Figure \ref{fig:Resid}. These residual maps are useful to assess goodness-of-fit. The colors indicate the significance ($\sigma$) of the residual (calculated as the difference between the data and the model) in each energy band used in this work. Positive residuals indicate regions that are underpredicted whereas negative ones indicate overpredicted regions. The means of the significance distribution shown in the residual maps of Figure  \ref{fig:Resid} are $-0.2 \pm 1.2$ (for the 60--300 MeV energy band); $0.1 \pm 2.5$ (300 MeV--3 Gev); $0.0 \pm 1.2$ (3--10 GeV) and $-0.1 \pm 1.1$ (10--500 GeV).

There are some regions with $|\sigma| > 5$ in the models with energies $>$300 MeV, especially in the 300 MeV--3 GeV energy band. They are the result of the minimal fitting we did when splitting the analysis in the different energy bands. This prevented us from getting better models but provided models that are comparable between the 3 highest-energy bands. To assess whether the results we obtained were, instead of a precise representation of the data, an artifact of the minimal fitting, we developed a simple test. We created unique models from scratch (instead of starting with the universal model) for each of the 3 highest-energy bands and performed the same analysis. In the tests, the results for each energy band were compatible within 1$\sigma$ with the ones presented in this work, both in terms of the $\gamma$-ray fluxes and the position of the source.

\begin{table*}
\centering
\begin{tabular}{c|ccccc}
\hline \hline
Model's&
\multicolumn{5}{c}{Spectral parameter}\\
\cline{2-6}
energy range&
$N_0$  & $E_0$ & \multirow{2}{*}{$\alpha$} & \multirow{2}{*}{$\beta$} & \multirow{2}{*}{$\Gamma$} \\
(GeV)& ($\times$10$^{-12}$ cm$^{-2}$ s$^{-1}$ MeV$^{-1}$) & (MeV) & & &
\\
\hline \hline
0.06--0.3 & 13.3 $\pm$ 2.7 & 2545 & - & - & -1.745 $\pm$ 0.073 \\
\hline

0.3--3 & 4.49 $\pm$ 0.09  & 4134 & 2.58 $\pm$ 0.03 & 0.234 $\pm$ 0.022 & - \\

3--10 & 2.52 $\pm$ 0.07 & 4134 & 2.79 $\pm$ 0.10 & 0.234 $\pm$ 0.021 & - \\

10--500 & 2.27 $\pm$ 0.06 & 4134 & 2.33 $\pm$ 0.06  & 0.234 $\pm$ 0.021 & - \\
\hline

0.1--500 & 2.53 $\pm$ 0.05 & 4134 & 2.59 $\pm$ 0.03 & 0.260 $\pm$ 0.011 & - \\
\hline
\end{tabular}
\caption{Spectral parameters obtained for 4FGL J1745.6$-$2859 after the fitting processes. We used a power law (Eq. \ref{Eq:PL}) for the 0.06--0.3 GeV energy range and a log-parabola (Eq. \ref{Eq:LogNorm}) for the other energy ranges.}
\label{TableSpectralParameters}
\end{table*}

\begin{figure*}
 \centering
 \begin{tabular}{@{}c@{}}
  \includegraphics[width=.49\linewidth]{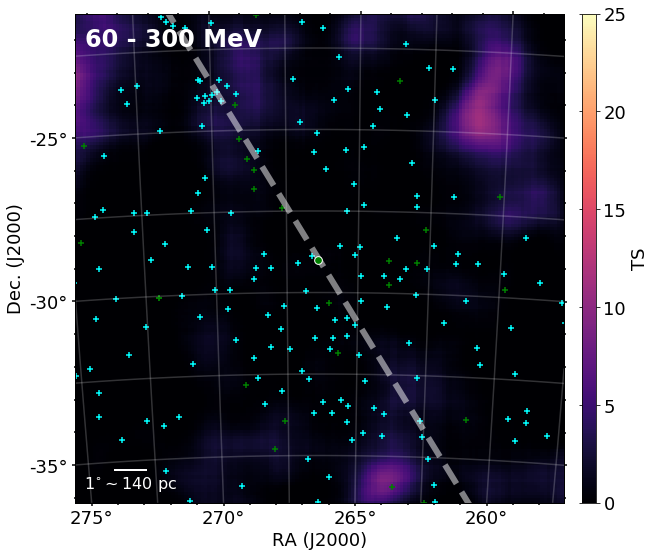}
  \hspace{0.02\linewidth}
  \includegraphics[width=.49\linewidth]{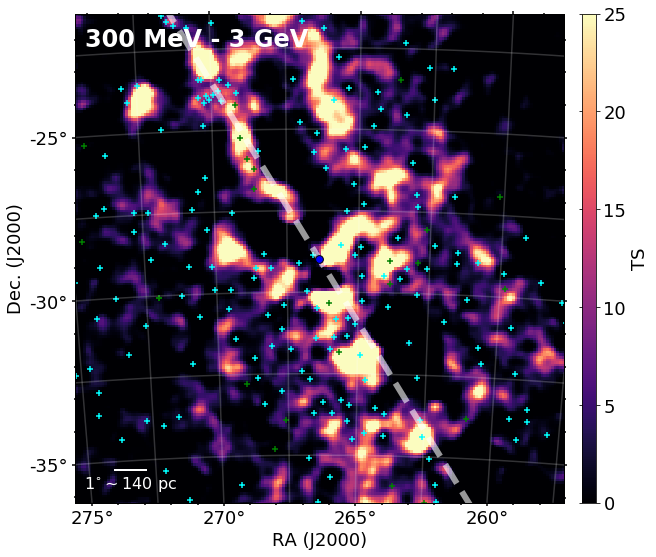} 
 \end{tabular} \\ 
 \begin{tabular}{@{}c@{}}
  \includegraphics[width=.49\linewidth]{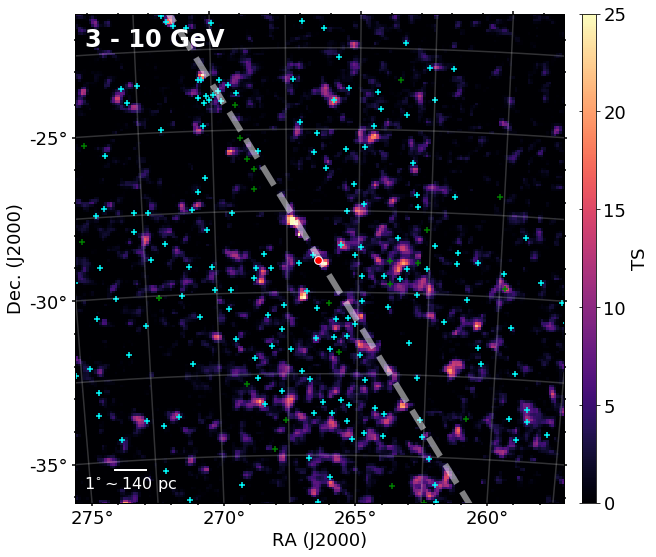}
  \hspace{0.02\linewidth}
  \includegraphics[width=.49\linewidth]{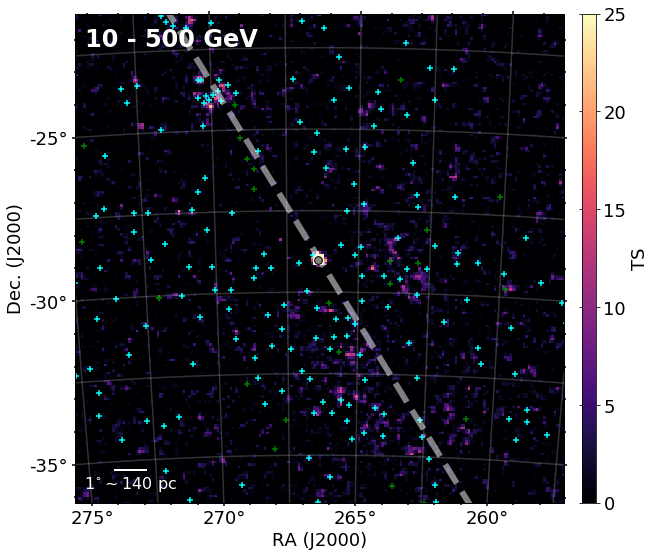}
 \end{tabular}
 \caption{TS maps of the RoI for the four different energy ranges. The central point source is not prominent in these maps since it is part of the model. The circles at the center of the panels correspond to the central point source position obtained in each energy range. 4FGL point sources are displayed as cyan crosses and new sources found during the analysis as green crosses. The gray dashed lines indicate the direction of the Galactic equator. An angular separation of 1$^{\circ}$ corresponds to $\sim$ 140 pc at \sgra's distance (8.2 kpc).}
\label{fig:TScom}
\end{figure*}

The four panels in Figure \ref{fig:TScom} show the TS map for each energy range. They were constructed using the \code{tsmap} tool. This tool moves a putative point source through the RoI and performs a maximum likelihood fit at each point. We used a power-law spectral model with a spectral index of $-2$ (with $dN/dE \propto E^{\alpha}$ where $dN/dE$ is the differential photon flux and $\alpha$ is the spectral index). In the maps, the source position in each energy range is marked by a colored circle in the center of the images. The central point source itself is not visible in the maps since it is part of the model. For the 60--300 MeV range there is no region with TS $\ge$ 25 (i.e. no emission consistent with a point source with a power-law spectrum with index $-2$ has significance  $\gtrsim 5\sigma$). This shows that the models include all significant sources in this field. But in the maps of the high-energy ($>$ 300 MeV) models this is not true. These models were created with minimal fitting based on the universal model (Section \ref{sec:obsHE}) which led to regions that could be better modeled if its sources were refit. Nevertheless, the results we got for these energy models are compatible within 1$\sigma$ with the ones obtained through models created specifically for each energy band (and that had no residual TS $> 25$).

The central source was detected in the four energy ranges used in the analysis with TS varying from $\approx300$ to $\approx 10000$, corresponding to detections with significance above background ranging from $\approx 17 \sigma$ to $\approx 100 \sigma$. Its photon and energy flux were also measured and the results are shown in Table \ref{Results}.

Assuming a distance to 4FGL J1745.6$-$2859 of 8.2 kpc \citep{Gravity2019} and isotropic emission, the energy flux of $(3.26 \pm 0.05) \times 10^{-10} \ {\rm erg \ cm^{-2} \ s^{-1}}$ measured in the 100 MeV to 500 GeV energy range corresponds to a $\gamma$-ray luminosity of $ (2.61 \pm 0.05) \times 10^{36} \ {\rm erg \ s}^{-1}$. This luminosity is comparable to the observed radio-to-X-ray luminosity of \sgra\ $\sim 10^{36} \ {\rm erg \ s^{-1}}$ \citep{Genzel10}.

The energy flux we measured in the 60--300 MeV is smaller than the results obtained by other works focused on previous versions of the cataloged GC $\gamma$-ray source: \cite{Chernyakova11} observed the \fermi-LAT source 1FGL J1745.6$-$2900 and \cite{Malyshev15} studied 2FGL J1745.6$-$2858. The energy fluxes we observe are about a factor of 2 lower than the ones reported by the former and only marginally consistent with the latter. This difference can be explained by two reasons. Firstly, in the following version of the \fermi\ Catalog, the 2FGL J1745.6-2858 source was split into two different sources: 3FGL J1745.6$-$2859c and 3FGL J1745.3$-$2903c. The former was subsequently renamed to 4FGL J1745.6$-$2859. \cite{Chernyakova11} and \cite{Malyshev15} were, then, considering the summed emission of what is now described as two different sources. The 3FGL J1745.3$-$2903c is softer than 3FGL J1745.6$-$2859c (and 4FGL J1745.6$-$2859), hence the greater difference between the  results in lower energies. Another reason to explain this difference is the improvement of the Galactic diffuse emission model through the years\footnote{https://fermi.gsfc.nasa.gov/ssc/data/analysis/software/aux/4fgl/\\Galactic$\_$Diffuse$\_$Emission$\_$Model$\_$for$\_$the$\_$4FGL$\_$Catalog$\_$Analysis.pdf}. In comparison to ours, their observations were potentially more contaminated with the diffuse emission associated with gas in the Galaxy, which is more prominent at lower energies.

\begin{figure*}
 \centering
 \begin{tabular}{@{}c@{}}
   \includegraphics[width=.49\linewidth]{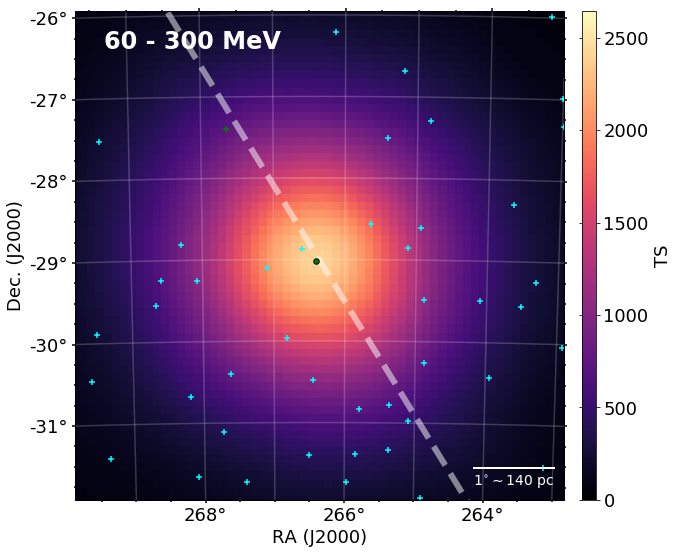}
   \hspace{0.02\linewidth}
   \includegraphics[width=.49\linewidth]{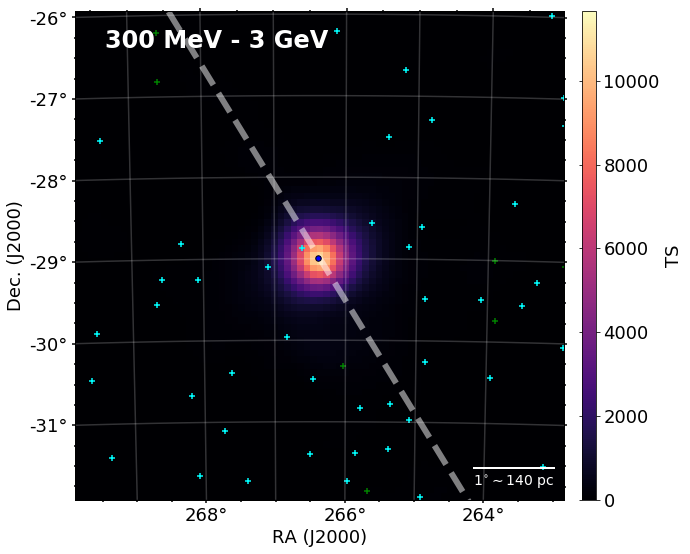} 
 \end{tabular} \\ 
 \begin{tabular}{@{}c@{}}
   \includegraphics[width=.49\linewidth]{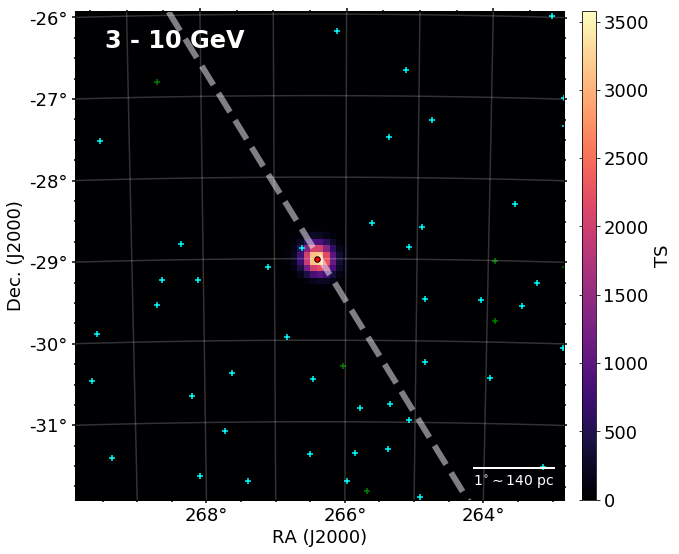}
   \hspace{0.02\linewidth}
   \includegraphics[width=.49\linewidth]{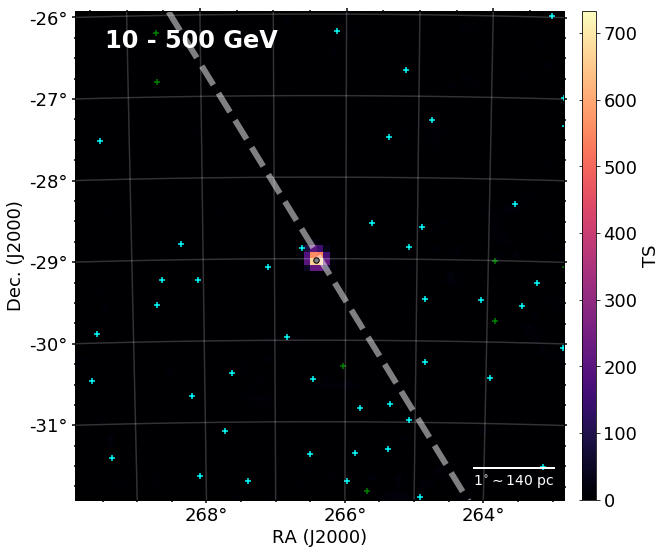}
 \end{tabular}
 \caption{TS maps of the inner $8^{\circ} \times 8^{\circ}$ of the RoI showing the contribution of the central point source. They were constructed after excluding 4FGL J1745.6$-$2859 from the models, as explained in the text. The point at the center of each panel corresponds to the source position obtained in each energy range. The other 4FGL point sources are shown as cyan crosses and new sources found during the analysis are show as green crosses. The apparent sizes of the excesses decrease with energy, but this does not mean any variation of the physical extension of the source. Instead, it is an outcome of the broadening of \fermi-LAT's PSF in low energies. The gray dashed lines indicate the direction of the Galactic equator. An angular separation of 1$^{\circ}$ corresponds to $\sim$ 140 pc at \sgra's distance (8.2 kpc).}
 \label{fig:TSsem}
\end{figure*}

The four panels of Figure \ref{fig:TSsem} show the TS maps of the inner $8^{\circ} \times 8^{\circ}$ of the RoI, showing the presence of the central point source. They were constructed using the \code{tsmap} tool (again considering a putative point with a spectral index of $-2$) and excluding the source 4FGL J1745.6$-$2859 from the models. TS excesses are observed in every panel and are always coincident with 4FGL J1745.6$-$2859's position. The apparent size of the excess in the TS maps seems to decrease with energy, but this does not correspond to any variation of the physical extension of the source. It is, instead,  an outcome of the improvement of \fermi-LAT's PSF with energy, as mentioned in Section \ref{sec:obs}.

The same diagnostic plots shown in Figures \ref{fig:Resid}, \ref{fig:TScom} and \ref{fig:TSsem} for the different energy bands were used to evaluate the quality of the fit of the universal model. They are shown in Appendix \ref{sec:UnivModelDiagnostic}.

\begin{figure}
\centering
\includegraphics[width=\linewidth]{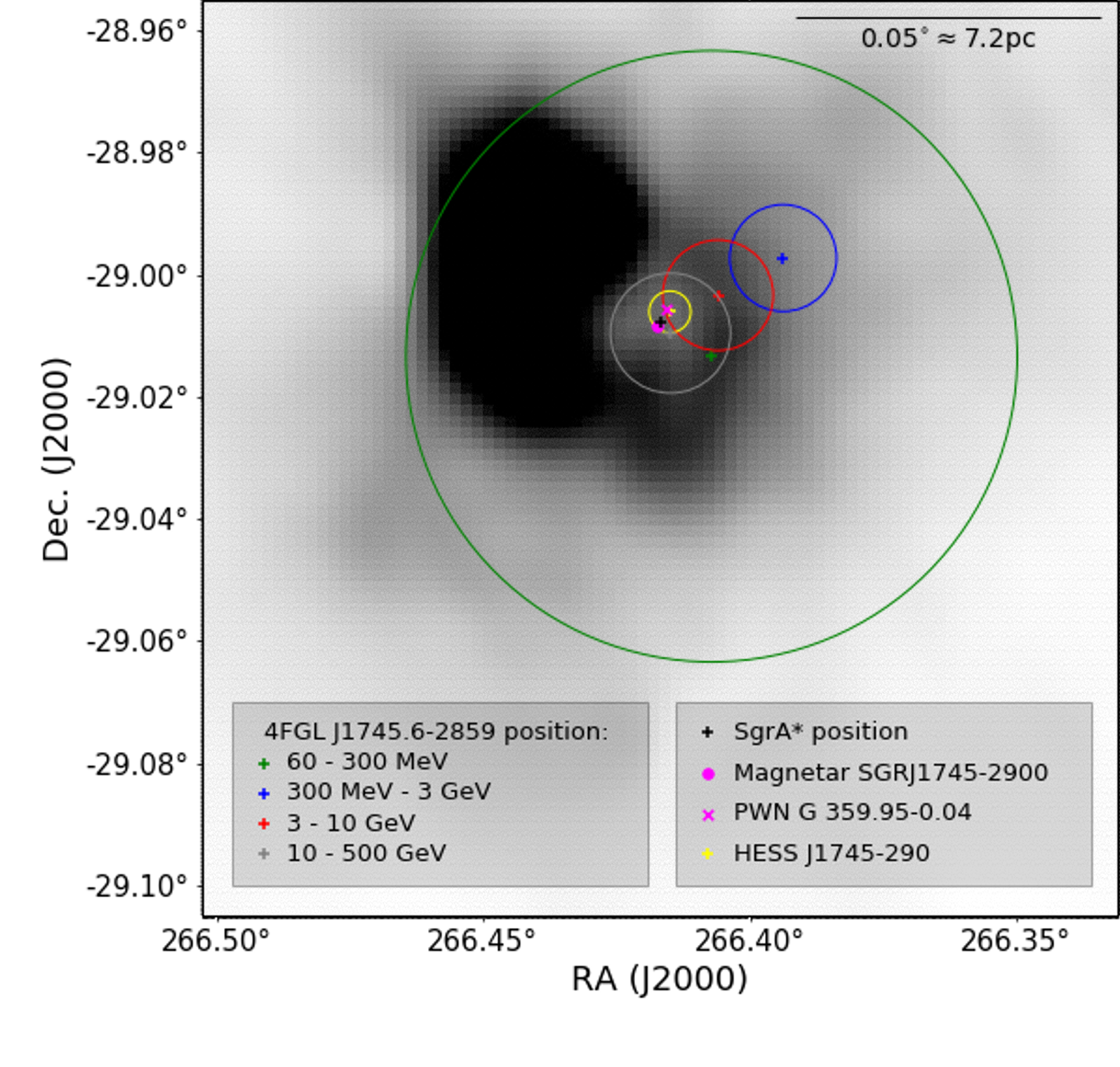}
\caption{The position of the central source as a function of the energy range used in the analysis: green (100--300 MeV), blue (300 MeV--3 GeV), red (3--10 GeV) and gray (10--500 GeV). The circles represent the 68$\%$ positional uncertainty. The radio position of \sgra\ is indicated by the black cross. The positions of other $\gamma$-ray-emitters in the GC are also indicated. The gray scale represents the 90 cm radio flux density map of Sagittarius A East \citep{LaRosa2000}, with darker color meaning brighter emission.}
\label{fig:points}
\end{figure}

We used the \code{localize} tool to constrain the point source location in each energy range. Figure \ref{fig:points} shows the dependence of the source location on energy, together with the radio position of \sgra\ as measured by the Very Long Baseline Array \citep{Petrov11} and locations of other potential $\gamma$-ray emitters in the GC.

To calculate the total errors $\Delta_{tot}$ on the location of the source we followed the approach used by \cite{4FGL} for the creation of the 4FGL Catalog:
\begin{equation}
    \Delta^2_{tot} = (f_{rel} \Delta_{stat})^2 + \Delta^2_{abs}
\label{eq:systematics}
\end{equation}

\begin{table*}
\centering
\begin{tabular}{cccccccc}
\hline \hline
Energy &
TS$^1$ &
Photon flux &
Energy flux &
Centroid$^2$ &
\multicolumn{2}{c}{Positional Uncertainty$^3$} &
Extension
\\
range (GeV) &
&
($\times$10$^{-7}$ cm$^{-2}$ s$^{-1}$) &
($\times$10$^{-10}$ erg cm$^{-2}$ s$^{-1}$) &
($^\circ$) &
statistical ($^\circ$) &
total ($^\circ$) &
UL$^4$ ($^\circ$) \\
\hline 

0.06 -- 0.3 & 2246 & 5.17 $\pm$ 0.16 & 1.06 $\pm$ 0.03 & 266.407, $-29.013$ & 0.045 & 0.050 & 0.24\\

0.3 -- 3  & 10522 & 1.49 $\pm$ 0.17 & 1.99 $\pm$ 0.20 & 266.394, $-28.997$ & 0.005 & 0.009 & 0.11\\

3 -- 10  & 3618 & 0.0894 $\pm$ 0.0022 & 0.669 $\pm$ 0.017 & 266.406, $-29.003$ & 0.005 & 0.009 & 0.08\\

10 -- 500  & 321 & 0.0123 $\pm$ 0.0011 & 0.345 $\pm$ 0.035 & 266.415, $-29.010$ & 0.005 & 0.010 & 0.05\\ \hline

0.1 -- 500  & 14724 & 2.83 $\pm$ 0.08 & 3.26 $\pm$ 0.05 & & &\\

\hline
\end{tabular}
\begin{tablenotes}\footnotesize
\item[1] $^1$ $\sqrt{TS} \approx$ detection significance of the source in each energy range
\item[2] $^2$ RA and Dec (in the J2000 epoch) corresponding to the emission centroid in degrees
\item[3] $^3$ 68\% confidence level positional uncertainty
\item[4] $^4$ 95\% confidence level extension upper limit
\end{tablenotes}
\caption{Results from the likelihood modeling of central point source. The last line presents the results for the universal model. The photon and energy flux uncertainties are statistical only.}
\label{Results}
\end{table*}

For the absolute precision $\Delta_{abs}$ we used the value of $0.^{\circ}0068$ \citep{4FGL} in the two energy bands below 3 GeV and $0.^{\circ}0075$ \citep{3FHL} for higher-energy bands. For the systematic factor $f_{rel}$ we used 1.1 for the three lower-energy bands and 1.2 for the 10--500 GeV band, these are conservative values based on \fermi-LAT's reported PSF systematic uncertainty\footnote{https://fermi.gsfc.nasa.gov/ssc/data/analysis/\\LAT$\_$caveats.html}. The statistical errors $\Delta_{stat}$ are the 68\% positional uncertainties obtained reported by the \code{localize} tool.

\begin{figure}
\centering
\includegraphics[width=\linewidth]{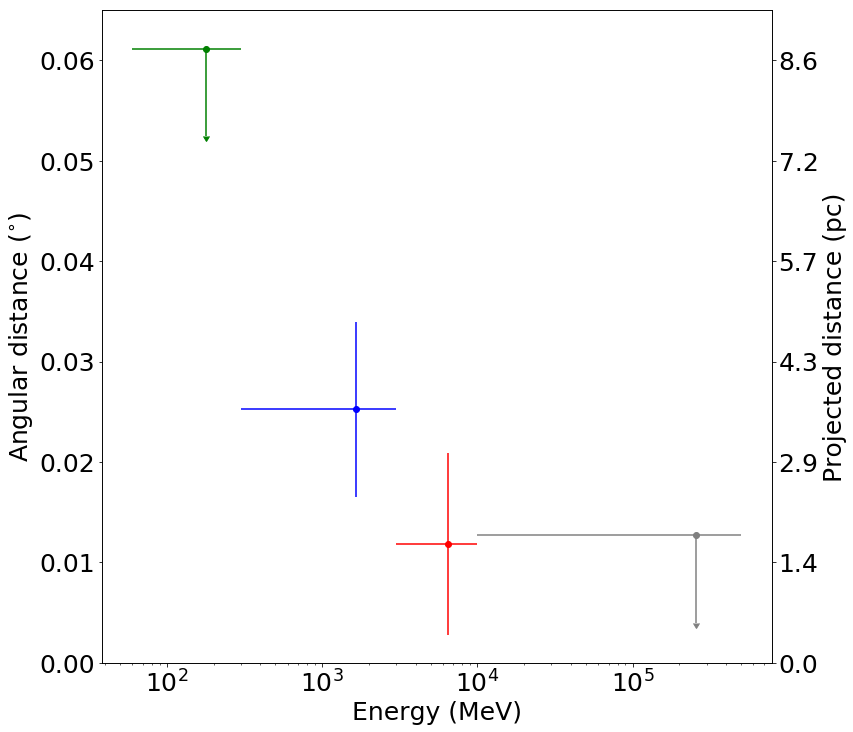}
\caption{The distance between the central source and \sgra\ as a function of energy. For the 100--300 MeV and 10--500 GeV energy ranges we show the 68$\%$ confidence level upper limits.}
\label{fig:distance}
\end{figure}

In Figure \ref{fig:points} we can see that the position uncertainty of the source in the lowest energy range is the largest. 
This is the result of a combination of factors: the broadening of \fermi-LAT's PSF at lower energies, the energy dependence of the instrument's field of view and effective area, the central source's spectrum and the impact of the Galactic diffuse emission which is more prominent at lower energies. The position of the source is consistent within $1\sigma$ with \sgra in the energy ranges 60--300 MeV and 10--500 GeV. The $\gamma$-centroid recedes from \sgra\ as the energy decreases. This is also seen in Figure \ref{fig:distance}, which shows the distance between 4FGL J1745.6$-$2859 and \sgra\ as a function of energy.

We also tested the likelihood of the source being extended versus a point source using the \code{extension} tool. Figure \ref{fig:extension} shows the 95\% upper limits on the spatial extension of the source for each energy range. We find no conclusive evidence for any spatially extended emission from the central source. 

\section{Discussion}\label{sec:disc}

By assuming a distance of 8.2 kpc for 4FGL J1745.6$-$\\2859 (i.e., that it is located at the GC) we obtain the interesting result that its $\gamma$-ray luminosity is similar to \sgra's radio to X-ray luminosity, about $10^{36} \ {\rm erg \ s^{-1}}$. 

\begin{figure}
\centering
\includegraphics[width=\linewidth]{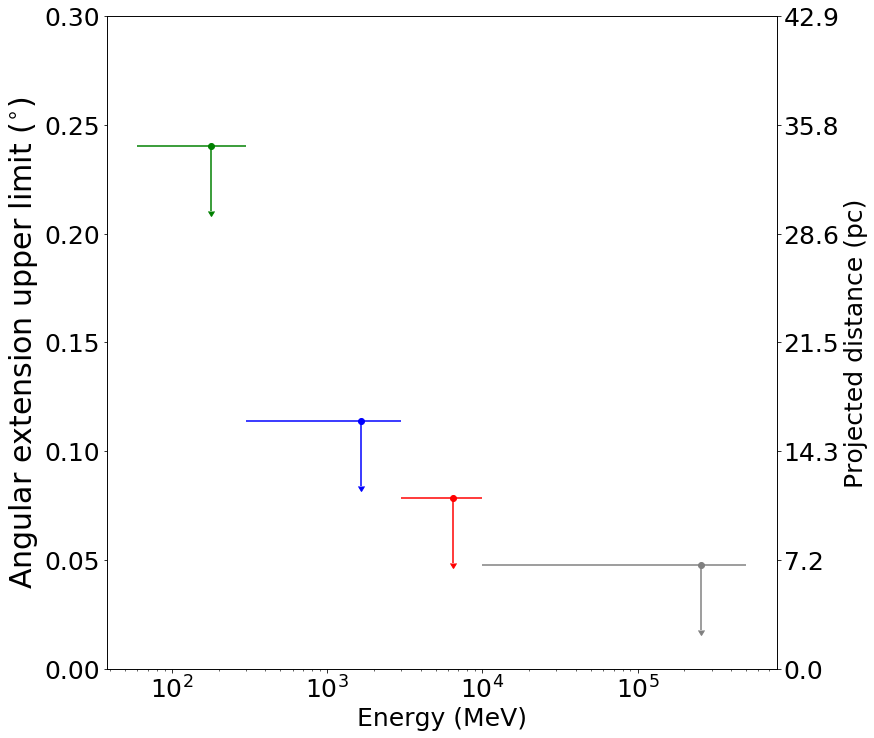}
\caption{Upper limit on the spatial extension of 4FGL J1745.6$-$2859 as a function of energy. Upper limits correspond to the 95$\%$ confidence level.}
\label{fig:extension}
\end{figure}

Even though this $\gamma$-ray luminosity is not unusual among \fermi\ sources\footnote{For instance, the point source 4FGL J1746.4$-$2852 lies just $\gtrsim 0.2^\circ$ from \sgra with a luminosity of $\sim 5.6 \times 10^{35} \ {\rm erg \ s^{-1}}$. It is associated with a pulsar wind nebula and was included in our source model.} the strong similarity between the electromagnetic energetics of 4FGL J1745.6$-$2859 and \sgra, combined with the positional coincidence, naturally suggests that the $\gamma$-ray point source investigated in this work is associated with the accreting SMBH. Nevertheless, the $0.1^\circ-1^\circ$ PSF of \fermi-LAT encompasses a region of size $\sim 10-100$ pc around the GC. Even the more constraining limit of $\lesssim$ 0$^{\circ}$.24 for the central source's extension upper limit (Figure \ref{fig:extension}) corresponds to $\lesssim$ 35 pc at the distance of \sgra, thus allowing for several other potential candidates for the $\gamma$-ray production site. Here, we list the most promising ones and discuss their likelihood at accounting for 4FGL J1745.6$-$2859.

\uline{The SMBH:} \cite{Aharonian05} argue that due to \sgra's low bolometric luminosity compared to other SBMHs, the $\gamma$-rays produced close to the event horizon---or in the inner parts of the accretion flow---can escape the source and be detected by \fermi-LAT because the absorption through photon-photon pair production is low. \cite{Aharonian05} considered three scenarios for the TeV photons detected by  H.E.S.S., two being hadronic and one leptonic. The first hadronic model considers emission related to accelerated protons producing $\gamma$-rays through synchrotron and curvature radiation. It predicts an energy flux lower than a few 10$^{-12}$ erg cm$^{-2}$ s$^{-1}$ in the energy range studied in this work, below the values we observed. The second hadronic scenario considers lower energy protons accelerated by the electric field close to the event horizon or by shocks in the accretion disks. Some parametrizations of this model predict very peaked spectral energy distributions (SEDs) in the energy ranges used in this work. Since these SEDs are very narrow, the energy fluxes they predict are consistent only with the observations in one of the four energy ranges we used in this work. Their leptonic model, in its turn, also fails to explain \fermi-LAT's observation of 4FGL J1745.6$-$2859: its SED shows $E^2 dN/dE \approx 4 \times 10^{-9}$ erg cm$^{-2}$ s$^{-1}$ at $\approx$1 GeV, thus overpredicting our observed energy flux around this energy.  On the other hand, \cite{Kusunose12} used 25 months of \fermi-LAT's data for the GC (reported by \citealt{Chernyakova11}) and proposed another leptonic model in which electrons escaping from the vicinity of \sgra\ accumulate in a region with a size of 10$^{18}$ cm where the $\gamma$-rays are produced by IC scattering of soft photons emitted by stars and dust around the GC. Importantly, they obtain energy fluxes similar ($\sim$10$^{-10}$ erg cm$^{-2}$ s$^{-1}$) to the values observed here.

\uline{A ``plerion'' produced by electrons in \sgra's winds:} \cite{Atoyan04} propose a model for \sgra\ in which the quiescent radio and the flaring NIR and X-ray emissions are generated by synchrotron radiation from the RIAF. The wind from the RIAF, in a process similar to pulsar-powered plerions, generates the quiescent X-ray and TeV emissions at the wind termination shock at about 3 $\times$ 10$^{16}$ cm from the SMBH. Although it can explain H.E.S.S.' TeV observations, their model is not sufficient to explain the MeV-GeV reported in this work. Even if we consider Sagittarius A West bremsstrahlung emission and the emission from a larger plerion (inflated to pc scales), which are prominent in the energy range used in this work (e.g., their Figure 1), the energy flux we detected is still about one order of magnitude higher. \cite{Atoyan04} assume that the quiescent X-ray emission detected from \sgra\ is the result of synchrotron radiation from electrons accelerated in the termination shocks of a wind of magnetized plasma from the accretion flow. But deep Chandra observations have shown that this emission is thermal and arises from the hot accretion flow \citep{Wang13}. Though these observations do not completely rule out the possibility of such a ``plerion'', this model now faces additional constraints from our \fermi-LAT observations.

\uline{The interaction between the dense molecular clouds with cosmic rays:} As an explanation for the $\gamma$-ray emission from the GC, \cite{Aharonian&Neronov05} presented a model of proton-proton interactions between the protons accelerated near the SMBH and the dense gas in the central 10 pc of the Galaxy which are followed by $\pi^{0}$ decay to $\gamma$-rays. Aharonian  \& Neronov's work was published before the beginning of operations of \fermi-LAT. Their results are inconsistent with our observations.

More recent models take into account \fermi-LAT observations. For instance, \cite{Chernyakova11} use the first 25 months of \fermi-LAT and H.E.S.S. data to create a hadronic model in which relativistic protons (presumably accelerated near \sgra) interact with the gas in the inner parsecs of the Galaxy. \cite{Linden12} developed a similar model. Another hadronic model is proposed by \cite{Fatuzzo12} where they consider a two-phase environment surrounding \sgra: an inner high-density ``torus'' and the surrounding interstellar medium filled with shocked stellar winds which they call the ``wind  zone''. \fermi\ $\gamma$-rays would be produced in the ``torus'' and the higher energies would come mostly from the ``wind zone''. \cite{Guo13} propose a hybrid model. In their scenario, protons and electrons are accelerated in the GC (possibly around \sgra). Collisions between the protons and the interstellar gas would produce the TeV $\gamma$-rays and the electrons would IC scatter the soft background photons. 

The four ``Fermi-era'' models mentioned above---namely, \citealt{Chernyakova11, Linden12, Fatuzzo12, Guo13}---are consistent with our observations, except in the lower energy band we used. This energy band is the one most subject to source confusion and to the impacts of the Galactic diffuse emission. This could explain why we observe greater energy fluxes in the 60--300 MeV energy band than the prediction of these models: in addition to the $\gamma$-rays created by the interaction between cosmic rays originated by (or nearby) \sgra, there is also a contribution from other sources. When \cite{Ahnen17} compared \fermi-LAT's data reported by \cite{Malyshev15} with the ``Fermi-era'' models, the lower energy data ($\lesssim$ 200 MeV) also showed greater fluxes than the models' predictions.

\uline{The PWN G 359.95-0.04:} This X-ray nebula was discovered by \cite{Wang06} with a projected distance of only 0.32 pc from \sgra\ and was proposed as an explanation for the TeV emission observed in the GC. \cite{Hinton07} constructed theoretical SEDs for this source based on Chandra's detection and supposing that the TeV emission of the H.E.S.S. source HESS J1745$-$290 is from the PWN. Their models (e.g., their Figure 4) predicts energy fluxes on the order of $\sim 10^{-12} \ {\rm erg \ cm^{-2} s}^{-1}$ for the energy ranges studied in this work. This is more than one order of magnitude lower than the energy fluxes we measured for 4FGL J1745.6$-$2859. PWN IC emission models are sensitive to the magnetic field intensity. \cite{Hinton07} adopted a magnetic field of $\sim$100 $\mu$G. Later measurements,  obtained thanks to the magnetar SGR J1745$-$2900 discovery, indicate that the GC magnetic field strength must be in the $\sim$mG levels \citep{Eatough13}. In more intense magnetic fields, synchrotron becomes more important as a path to electron cooling than IC. It means that if the \cite{Hinton07} model was updated with the newer observational data, its $\gamma$-ray output would be even more displaced from our observations. This indicates that G 359.95-0.04 is not a good candidate to explain 4FGL J1745.6$-$2859's emission.

\uline{The SNR Sagittarius A East:} Sagittarius A East is an SNR \sout{usually explained as a supernova remnant, although other interpretations have also been suggested} \citep{Yusef-Zadeh1987, Mezger1989, Khokhlov1996}. \sout{It is} located in the inner parsecs of the Galaxy. \cite{Crocker05} proposed it as the source of the TeV $\gamma$-rays from the GC. On the other hand, \cite{Aharonian09} and \cite{Acero10} ruled out this association based on H.E.S.S. observations that show the origin of the TeV emission, although still coincident with Sagittarius A East's extended radio shell, coming from a region where the radio flux is comparatively low and significantly displaced from the radio maximum, as indicated in Figure \ref{fig:points}. The same argument can be used to rule out a physical link between the SNR and the 4FGL J1745.6$-$2859 in energies above 300 MeV.

\uline{The magnetar SGR J1745$-$2900:} This object was first detected during a flare in 2013 with  \textit{Swift}'s X-Ray Telescope \citep{Kennea13} and NuSTAR \citep{Mori13}. 4FGL J1745.6$-$2859's $\gamma$-ray light curve shows no sign of variability during this period (\citealt{Malyshev15, Ahnen17}; Cafardo et al., in preparation). Also, it is predicted that the high-energy portion of the spectra of magnetars peak at a few MeV \citep{Thompson05}, while our work clearly shows emission from 4FGL J1745.6$-$2859 at energies $>$ 10 GeV.

\uline{Self-annihilating dark matter particles accumulating at the GC:} Self-annihilating dark matter particles could explain the GC $\gamma$-ray excess (\citealt{Hooper11, Hooper&Linden11} and see \citealt{DiMauro21} for a recent analysis of the GC excess), a surplus of $\sim$ GeV diffuse emission that cannot be explained by the known cataloged sources. Although we did not test different morphologies for 4FGL J1745.6$-$2859, we did investigate the likelihood of it being extended, which is expected for the $\gamma$-ray from self-annihilating dark matter. Using a 2D Gaussian profile, we found no evidence of extension, hence decay profiles with similar morphologies are disfavored.

\uline{A population of pulsars surrounding the GC:} The $\gamma$-ray spectra of pulsars and millisecond pulsars can be described as a power-law with an exponential cutoff above a few GeV \citep{Abdo10_pulsar}. \cite{de_Menezes19} studied the $\gamma$-ray emission of globular clusters in the Milky Way---attributed to their large population of millisecond pulsars---and found no significant flux above $\sim 10$ GeV. In contrast with that, the point source 4FGL J1745.6$-$2859 is detected at energies above that as indicated in Table \ref{Results} and Figure \ref{fig:TSsem}. Its hard spectrum is also not consistent with a pulsar (or population of them) in the line of sight. On the other hand, \cite{Bednarek13} propose that the GeV $\gamma$-ray emission from the GC is the result of the millisecond pulsar population of several globular clusters captured by \sgra, while the TeV portion is produced by accelerated leptons in their winds via IC process in the soft radiation field. Since the TeV component is harder, it can start to contribute above $\sim$10 GeV in their model, which could be compatible with our observations. A detailed spectral analysis of 4FGL J1745.6$-$2859 is necessary to test whether \cite{Bednarek13} model is compatible with the GC $\gamma$-ray source.

\uline{The central cluster of massive stars:} A young and dense stellar cluster lies in the GC \citep{Do2013}. \cite{Quataert05} propose a model in which shocks from stellar winds can efficiently accelerate electrons to relativistic energies. Then, they IC scatter the ambient radiation field producing $\gamma$-rays from $\sim$GeV to $\sim$10 TeV. Considering a distance of 8.2 kpc to the GC, their model predicts energy flux one order of magnitude lower than our observations. \cite{Aharonian19} have discussed young stellar clusters as sources of cosmic rays. They suggest that the cosmic rays responsible for the diffuse very high energy $\gamma$-ray form the GC are accelerated by the local stellar clusters. This emission, though, is extended while we find no evidence for 4FGL J1745.6$-$2859.
\newline

Except for the models constructed around \fermi-LAT's data, most of the candidates listed above are unlikely to be solely responsible (if responsible at all) for 4FGL J1745.6$-$2859's emission. It is possible that some of them could explain, individually or together, the lower energy emission where \fermi-LAT's PSF is broadest. One way to separate the contributions of different candidates at lower energies is through modeling the MeV-to-GeV SED of the point source. This is beyond the scope of the present paper, and left for a forthcoming work.

If we consider only the three highest-energy ranges used in this analysis, the centroid emission moves in the direction of \sgra\ as the photon energy is increased (Figures \ref{fig:points} and \ref{fig:distance}). Assuming a distance of 8.2 kpc, the projected distances to \sgra\ as a function of energy varies from 3.6 $\pm$ 1.3 pc (300 MeV--3 GeV) to 0.4 $\pm1.4$ pc (10--500 GeV). The location centroid for the lower-energy band is also consistent within 1$\sigma$ with \sgra's. The absence of a significant offset of the centroids with \sgra---and with each other---is consistent with an origin close to the SMBH. One possible interpretation is that the particle populations responsible for the $\gamma$-rays detected in the three bands between 300 MeV and 500 GeV are accelerated by the same process, originating in the surroundings of \sgra.

As discussed in Section \ref{sec:obs}, the impact of source confusion on \fermi-LAT observations is greater at lower energies due to the PSF broadening. The localization uncertainties are considerably larger at low energies, as well as the limit on the angular extension, which creates the possibility that several other sources and processes are contributing to the lower-energy flux.

We performed extension analysis of the source in the four energy bands used in this work. In Figure \ref{fig:extension}, we report the 95\% confidence level upper limit for the spatial extension of the source. Again, we observe the impact of energy on the results: at lower energies the spatial extent is less well constrained due to the increasing PSF.

The association between 4FGL J1745.6$-$2859 and the GC's TeV source, although likely, is not yet established. Assuming they are not related, the hard spectrum of the TeV source is expected to contribute in the highest energy band used in this work. This would make the \fermi\ source's spectrum to appear harder than it really is, preventing the elimination of emitting candidates based on the 10--500 GeV energy range results. This will be investigated in a future work.

\section{Summary}	\label{sec:summary}

\sgra---the accreting SMBH at the center of our galaxy---has been observed in virtually every band of the electromagnetic spectrum. In $\gamma$-rays, H.E.S.S. and \fermi-LAT detected point sources coincident with \sgra. Nevertheless the connection between these point sources and \sgra\ remained inconclusive. In this work we have used about 10.5 years of \fermi-LAT observations of the point source 4FGL J1745.6$-$2859 at the Galactic Center with the aim of constraining the nature of its emission. We divided the analysis into four different energy bands between 60 MeV and 500 GeV, performing a detailed imaging analysis of the surroundings of the point source. Our main conclusions can be summarized as follows:

(i) The 0.1--500 GeV luminosity of the point source is $ (2.61 \pm 0.05) \times 10^{36} \ {\rm erg \ s}^{-1}$ assuming it is located at the Galactic Center; this value is comparable to the observed luminosity of \sgra\ from radio to X-rays.

(ii) The point source location approaches \sgra's position as the photon energy is increased. For instance, at energies $> 10$ GeV the source location is consistent with \sgra\ within $1\sigma$.

(iii) Among several possible candidates to the $\gamma$-ray flux of the point source, the models invoking cosmic rays---either hadronic or leptonic---accelerated by \sgra\ or a nearby source are the most likely to explain our observations.

(iv) Other processes not associated with the SMBH, such as a population of pulsars in the GC, could be contributing to the flux, especially at energies $<$ 300 MeV.

Taken together, our results support the picture in which the point source 4FGL J1745.6$-$2859 observed by \fermi-LAT at the GC is the manifestation of \sgra\ in the MeV-to-GeV range. 

The advent of the Cherenkov Telescope Array (CTA, \citealt{CTA_2019}) will allow for a deep exposure of the GC in energies up to $\sim$300 TeV. This will permit studies in spatial and spectral details unavailable today, with arc-minute resolution at energies above \fermi's operational range, potentially enabling a firmer association between the very high-energy point source in the GC with \sgra\ or other nearby candidate. Correspondingly, proposed $\gamma$-ray missions focusing on MeV bands such as the AMEGO mission \citep{AMEGO} and e-ASTROGAM \citep{e-ASTROGAM} should improve the observational sensitivity, helping to better constrain the properties of the GC emission in the 60--300 MeV energy band and shed light on the contribution of \sgra.
\newline

In a forthcoming work, we will analyze the $\gamma$-ray variability and SED of 4FGL J1745.6$-$2859 in order to further constrain its physical origin.

\section*{Acknowledgments}	\label{sec:acknowledgments}

We thank the anonymous referee for their questions, comments and suggestions that greatly improved the quality of this work. We acknowledge useful discussions with John W. Hewitt, Raniere de Menezes, Teddy Cheung, Matthew Kerr, Regina Caputo, Aion Viana, Julia C. Santos, Giacomo Principe, Jeremy S. Perkins, Seth Digel and Pablo Araya-Araya. This work was supported by CNPq (Conselho Nacional de Desenvolvimento Cient\'ifico e Tecnol\'ogico) under grant 142320/2016-1 and FAPESP (Funda\c{c}\~ao de Amparo \`a Pesquisa do Estado de S\~ao Paulo) under grant 2017/01461-2. 

The \fermi-LAT Collaboration acknowledges generous ongoing support from a number of agencies and institutes that have supported both the development and the operation of the LAT as well as scientific data analysis. These include the National Aeronautics and Space Administration and the Department of Energy in the United States, the Commissariat \`a l'Energie Atomique and the Centre National de la Recherche Scientifique / Institut National de Physique Nucl\'eaire et de Physique des Particules in France, the Agenzia Spaziale Italiana and the Istituto Nazionale di Fisica Nucleare in Italy, the Ministry of Education, Culture, Sports, Science and Technology (MEXT), High Energy Accelerator Research Organization (KEK) and Japan Aerospace Exploration Agency (JAXA) in Japan, and the K.~A.~Wallenberg Foundation, the Swedish Research Council and the Swedish National Space Board in Sweden.
 
Additional support for science analysis during the operations phase is gratefully acknowledged from the Istituto Nazionale di Astrofisica in Italy and the Centre National d'\'Etudes Spatiales in France. This work performed in part under DOE Contract DE-AC02-76SF00515.

\facilities{\fermi-LAT}
\software{Fermipy \citep{wood2017fermipy}, APLpy \citep{aplpy2012, aplpy2019}.}

\clearpage

\appendix

\section{New sources found in the analysis between 100 MeV and 500 GeV}
\label{sec:newsources}

Here we present the new sources encountered in the RoI with \code{Fermipy}'s \code{find\_sources} function  in the energy range between 100 MeV and 500 GeV. The maximum likelihood parameters for the power-law (Equation \ref{Eq:PL}) used to model their spectra are also shown, together with their TS and position.

Some of these newly detected sources might be spurious due to unmodeled or inadequately modeled background emission.

\begin{table}[h]
\centering{
\begin{tabular}{cccccc}
\hline \hline
Source name &   Index   &   Prefactor   &  TS  &   RA  &   Dec\\
\hline
PS J1719.1$-$2945$^{\dagger}$     &   $-5.0$    &   2.8$\times$10$^{-14}$           &   236 &   259.80  &   $-29.75$\\
PS J1720.6$-$2655     &   $-2.1$    &   4.5$\times$10$^{-13}$           &   40  &   260.16  &   $-26.93$\\
PS J1723.8$-$3347     &   $-2.0$    &   5.8$\times$10$^{-13}$          &   31  &   260.95  &   $-33.79$\\
PS J1729.5$-$3623     &   $-1.9$    &   4.7$\times$10$^{-13}$           &   40  &   262.39  &   $-36.40$\\
PS J1730.5$-$2801     &   $-2.0$    &   3.8$\times$10$^{-13}$           &   31  &   262.64  &   $-28.03$\\
PS J1731.6$-$2903     &   $-1.8$	&   2.9$\times$10$^{-13}$	    	&   34	&   262.92	&   $-29.05$\\
PS J1733.8$-$2114	    &   $-2.0$	&   2.8$\times$10$^{-13}$	    	&   30	&   263.47	&   $-21.24$\\
PS J1734.4$-$3555	    &   $-2.2$	&   5.8$\times$10$^{-13}$	    	&   34	&   263.62	&   $-35.93$\\
PS J1734.6$-$2328	    &   $-2.2$	&   4.9$\times$10$^{-13}$	    	&   37	&   263.66	&   $-23.48$\\
PS J1735.5$-$2944	    &   $-1.8$	&   3.3$\times$10$^{-13}$	    	&   31	&   263.88	&   $-29.74$\\
PS J1735.6$-$2900$^{\dagger}$	    &   $-5.0$	&   1.6$\times$10$^{-14}$	    	&   36	&   263.91	&   $-29.01$\\
PS J1742.7$-$3150	    &   $-2.0$	&   8.0$\times$10$^{-13}$	    	&   50	&   265.68	&   $-31.85$\\
PS J1744.1$-$3019	    &   $-2.2$	&   2.7$\times$10$^{-12}$	    	&   124	&   266.03	&   $-30.32$\\
PS J1747.2$-$2114	    &   $-2.3$	&   5.8$\times$10$^{-13}$	    	&   40	&   266.81	&   $-21.25$\\
PS J1750.4$-$3355	    &   $-2.2$	&   5.6$\times$10$^{-13}$	    	&   40	&   267.61	&   $-33.93$\\
PS J1752.0$-$3447	    &   $-2.2$	&   4.1$\times$10$^{-13}$	    	&   34	&   268.02	&   $-34.79$\\
PS J1752.6$-$2105$^{\dagger}$	    &   $-5.0$	&   4.0$\times$10$^{-14}$	    	&   303	&   268.15	&   $-21.09$\\
PS J1754.4$-$2612	    &   $-2.1$	&   7.6$\times$10$^{-13}$	    	&   32	&   268.60	&   $-26.21$\\
PS J1754.4$-$2649	    &   $-2.0$	&   5.8$\times$10$^{-13}$	    	&   30	&   268.60	&   $-26.82$\\
PS J1754.7$-$3730	    &   $-2.2$	&   3.7$\times$10$^{-13}$	    	&   38	&   268.70	&   $-37.51$\\
PS J1755.3$-$2553	    &   $-2.0$	&   9.3$\times$10$^{-13}$	    	&   49	&   268.83	&   $-25.90$\\
PS J1756.0$-$3248	    &   $-2.1$	&   4.1$\times$10$^{-13}$	    	&   31	&   269.02	&   $-32.81$\\
PS J1756.3$-$2515	    &   $-2.3$	&   2.1$\times$10$^{-12}$	    	&   99	&   269.08	&   $-25.25$\\
PS J1756.8$-$2413	    &   $-1.9$	&   5.4$\times$10$^{-13}$	    	&   36	&   269.21	&   $-24.22$\\
PS J1758.0$-$2421	    &   $-2.2$	&   1.1$\times$10$^{-12}$	    	&   38	&   269.52	&   $-24.36$\\
PS J1808.2$-$3003	    &   $-2.2$	&   4.5$\times$10$^{-13}$	    	&   45	&   272.06	&   $-30.05$\\
PS J1817.4$-$2516	    &   $-2.4$    &   5.7$\times$10$^{-13}$       	&   50	&   274.35  &   $-25.27$\\
PS J1818.6$-$2812     &   $-2.6$	&   5.0$\times$10$^{-13}$       	&   57	&   274.65	&   $-28.22$\\
PS J1823.9$-$2341	    &   $-2.4$	&   5.1$\times$10$^{-13}$       	&   47	&   275.99	&   $-23.69$\\
\hline
\\
\end{tabular}
\caption{New sources found in the 100 MeV to 500 GeV energy range. RA and Dec are in the J2000 epoch. Sources marked with $^{\dagger}$ are very likely spurious.}
\label{TableNewSources}
}
\end{table}

\clearpage

\section{New sources found in the analysis with the custom model between 60 and 300 MeV}
\label{sec:newsorcesLowEnergy}

Five new sources were encountered in the RoI with \code{Fermipy}'s \code{find\_sources} function in the energy range between 60 and 300 MeV. They are listed in Table \ref{TableNewSourcesLowEnergy}. The best parameters for the power-law (Equation \ref{Eq:LogNorm}) used to model their spectra are also shown, together with their TS and position.

These might be spurious findings due to imperfections in the model.

\begin{table}[h]
\centering{
\begin{tabular}{cccccc}
\hline \hline
Source name &   Index   &   Prefactor     &   TS  &   RA  &   Dec\\
\hline
PS J1639.5$-$2448     &     $-1.7$     &     4.4$\times$10$^{-12}$         &     37     &     249.90     &     $-24.81$\\
PS J1750.6$-$2723     &     $-2.0$     &     4.8$\times$10$^{-12}$         &     27     &     267.65     &     $-27.40$\\
PS J1753.7$-$2127     &     $-0.9$     &     2.9$\times$10$^{-12}$         &     57     &     268.44     &     $-21.47$\\
PS J1820.5$-$2113     &     $-2.0$     &     4.3$\times$10$^{-12}$         &     61     &     275.14     &     $-21.22$\\
PS J1835.0$-$1804$^{\dagger}$     &     $-5.0$     &     8.8$\times$10$^{-15}$         &     183    &     278.76     &     $-18.07$\\
\hline
\\
\end{tabular}
\caption{New sources found in the 60 to 300 MeV energy range. RA and Dec are in the J2000 epoch. The source marked with $^{\dagger}$ is very likely spurious.}
\label{TableNewSourcesLowEnergy}
}
\end{table}

\clearpage

\section{Diagnostic plots for the universal model}
\label{sec:UnivModelDiagnostic}

We present the residual map, the TS maps with and without 4FGL J1745.6$-$2859 in the universal model created in the 100 MeV to 500 GeV energy range.

\begin{figure}[h]
 \centering
 \begin{tabular}{@{}c@{}}
   \includegraphics[width=.40\linewidth]{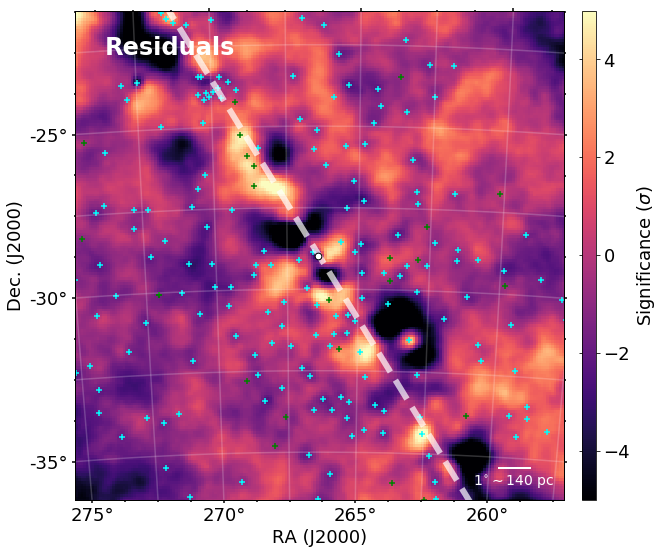}
   \hspace{0.01\linewidth}
 \end{tabular} \\ 
 \begin{tabular}{@{}c@{}}       \includegraphics[width=.40\linewidth]{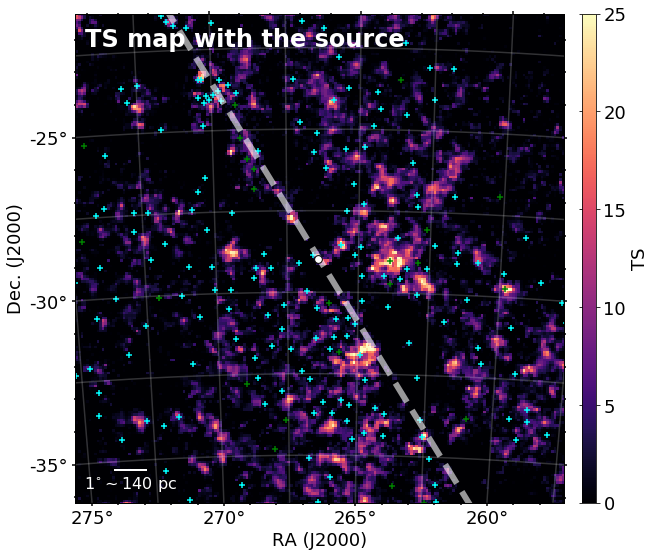}
   \hspace{0.01\linewidth}
    \end{tabular} \\ 
 \begin{tabular}{@{}c@{}}   \includegraphics[width=.41\linewidth]{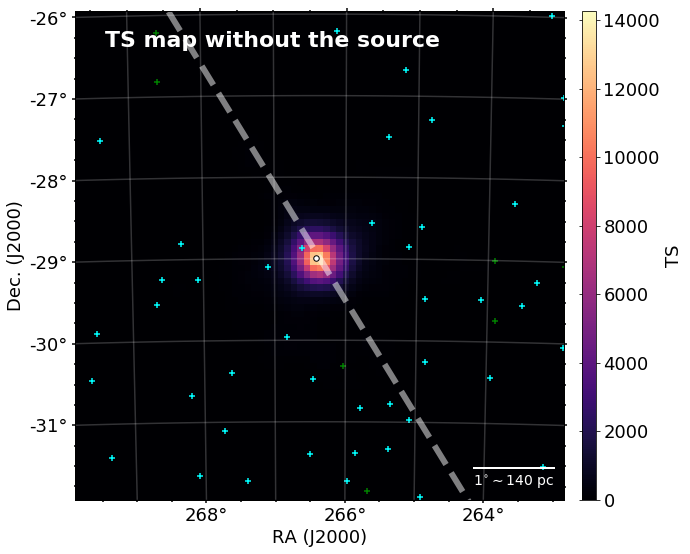}
   \hspace{0.01\linewidth}
 \end{tabular} \\ 
 \caption{Diagnostic plots for the universal model. In the upper panel, the colors show the significance of the residual (this distribution has mean $-0.5 \pm 2.2$). On the other panels, the colors indicate the TS value in each position. The point at the center of each panel corresponds to 4FGL J1745.6$-$2859 position in the 4FGL Catalog. 4FGL point sources are displayed as cyan crosses and new sources found during the analysis as green crosses. The gray dashed lines indicate the direction of the Galactic equator. An angular separation of 1$^{\circ}$ corresponds to $\sim$ 140 pc at \sgra's distance (8.2 kpc).}
 \label{fig:UnivModelDiagnostic}
\end{figure}

\bibliography{refs,refsNemmen}

\end{document}

%% file: definitions.tex
\usepackage[shadow]{todonotes}	
\usepackage{color, soul}	
\usepackage{wrapfig}	
\usepackage{verbatim}	
\usepackage[normalem]{ulem} 
\usepackage{tabularx} 
\usepackage[flushleft]{threeparttable} 

\newcommand{\fermi}{\emph{Fermi}}

\newcommand{\sgra}{Sgr A$^*$}

\newcommand{\orcid}[1]{\href{#1}{\includegraphics[scale=0.035]{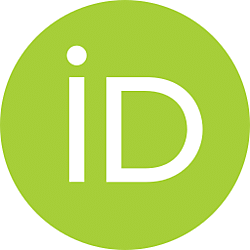}}}

\definecolor{boxback}{HTML}{DAE5F0}
\definecolor{boxframe}{HTML}{0F365B}
{
\end{tcolorbox}   
}

\newcommand{\eg}[1]{(e.g. \citealt{#1})}

\def\code#1{\texttt{#1}}

\def\say#1{\textbf{[#1]}}

